\documentclass[runningheads]{svmult}

\usepackage{makeidx}   
\usepackage{graphicx}  
\usepackage{subeqnar}  
\usepackage{multicol}  
\usepackage{physprbb}  
\makeindex             

\newcommand{\greeksym}[1]{{\usefont{U}{psy}{m}{n}#1}}

\newcommand{\uLambda}{\mbox{\greeksym{L}}}

\begin{document}
\title*{Galaxy Masses: Disks and Their Halos}
\toctitle{Galaxy Masses: Disks and Their Halos}
%
%
\titlerunning{Disk and Halo Masses}
%
\author{Stacy McGaugh}
\authorrunning{ McGaugh }
%
%
\institute{Department of Astronomy, University of Maryland,
College Park, MD 20742}

\maketitle              

\begin{abstract}
I review what we currently do and do not know about the masses of disk
galaxies and their dark matter halos.  The prognosis for 
disks is good: the asymptotic rotation velocity provides a good
indicator of total disk mass.  The prognosis for halos is bad:
cuspy halos provide a poor description of the data, and
the total mass of individual dark matter halos remains ill-constrained.
\end{abstract}

\section{Disk Masses}

The great regularity of the Tully-Fisher relation \cite{TF} has long been
thought to originate from a strong mass-velocity relation and 
a near constancy of mass-to-light ratio.  The latter requires a fair but
not unreasonable amount of regularity in stellar populations. 
Put simply,\footnote{``Stacy, you're a genius! ...[!] ...when it comes to
pepper grinders'' (van den Bosch 2001, private communication).}
\begin{equation}
L \sim {\cal M} \sim V^a \; .
\end{equation}

There have
long been indications (Sancisi 1995, private communication) that this
simple scaling may fail at low luminosities.  This has
become more clear as data have improved \cite{MG},\cite{Stil}.  This
breakdown of the Tully-Fisher relation 
might arise because of the chaotic star formation histories of low mass 
galaxies, or as a result of a breakdown in the underlying mass-velocity
relation.  Another possibility is that optical luminosity ceases to trace
mass because stars cease to be the dominant mass component in these disks
\cite{Ken}.

It has now become clear that this last possibility is in fact the case. 
Low mass galaxies
are often dominated by gas rather than stars.  If instead of luminosity
or stellar mass, we plot disk (star + gas) mass against the flat
rotation velocity, a nice mass-velocity
relation is recovered over many orders of magnitude (Fig.~1).
This `Baryonic Tully-Fisher Relation' (BTF) is \cite{MSBB}
\begin{equation}
{\cal M}_d = {\cal A} V_f^b \; ,
\end{equation}
for which the data in Fig.~1 give
\begin{subeqnarray}
{\cal A} = 50\;{\cal M}_{\odot}\;{\rm km}^{-4}\;{\rm s}^4 \nonumber \\
b =  4.0 \pm 0.1 \; .
\setcounter{eqsubcnt}{0}
\end{subeqnarray}

\begin{figure}[t]
\begin{center}
\includegraphics[width=.9\textwidth]{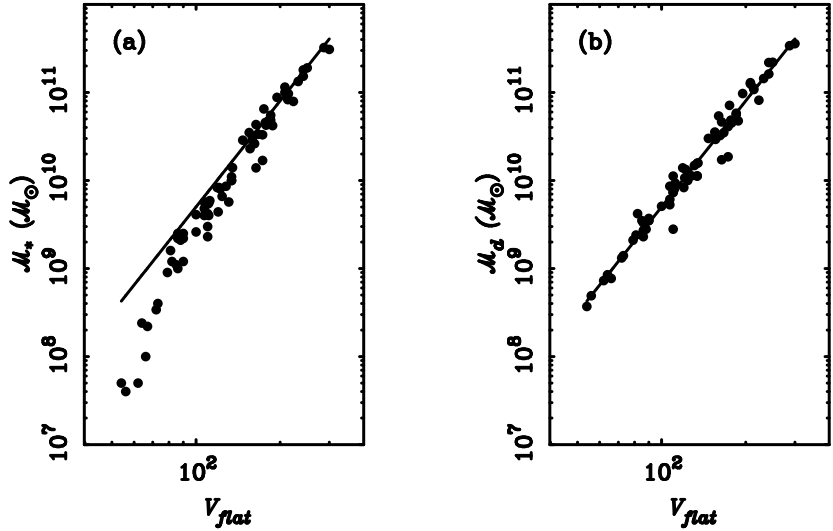}
\end{center}
\caption[]{The Tully-Fisher relation expressed in terms of
(\textbf{a}) stellar mass and (\textbf{b}) baryonic disk mass
(the sum of stars and gas).  The luminous Tully-Fisher relation holds
well for galaxies dominated by stars, but breaks down for low mass
galaxies where the gas mass can often exceed the stellar mass (\textbf{a}).
The sum of stars and gas provides a better correlation (\textbf{b}):
the asymptotic flat rotation velocity is a good indicator of disk mass
\cite{MSBB}.  The data shown here are taken from a large compilation of
high quality data \cite{SM}.  Consequently, the scatter is greatly
reduced from that in \cite{MSBB}.  The galaxies shown here are drawn from
the full range of disk Hubble types: mostly Sc, Sd, Sm, Im, but also a few
Sa and Sb galaxies are present.  The intrinsic scatter is small, with room
only for scatter in the stellar mass-to-light ratio due to variations in
star formation histories (probably not in the IMF), and scatter due to
the modest ellipticities of disks \cite{Bersh}.
}
\label{fig1}
\end{figure}

The normalization of the BTF is rather uncertain:  formally acceptable
values fall in the range $34 < {\cal A} < 85$.
The precise value of the slope has been modestly controversial:
$b= 4.0$ was given by \cite{MSBB} while $b=3.5$ was found by
\cite{BJ}.  This difference can be traced to different assumptions
about the (rather goofy \cite{HSTKP}) distance to the UMa cluster
for which some of the better rotation curve \cite{VS} and
photometric data \cite{TVPHW} exist.  As the distance increases, the gas mass
increases faster than the stellar mass (as $D^2$ and as $D$, respectively).
This boosts the total mass of gas dominated galaxies by a larger factor
than star dominated galaxies.  Since these reside at opposite ends of
the relation, the slope tips to shallower values with increasing $D$.
Nevertheless, the population models of \cite{BJ} are consistent
with a slope of $b= 4.0$ (Fig.~2).
While the calibration of the BTF can always be improved,
it already provides an excellent indicator of disk mass.
Moreover, continuity between gas-rich and star-rich galaxies constrains
stellar population mass-to-light ratios.  The favored
values are reasonable in terms of population synthesis models (Fig.~2),
but unpleasantly heavy for cuspy dark matter halos.

\begin{figure}[t]
\begin{center}
\includegraphics[width=.9\textwidth]{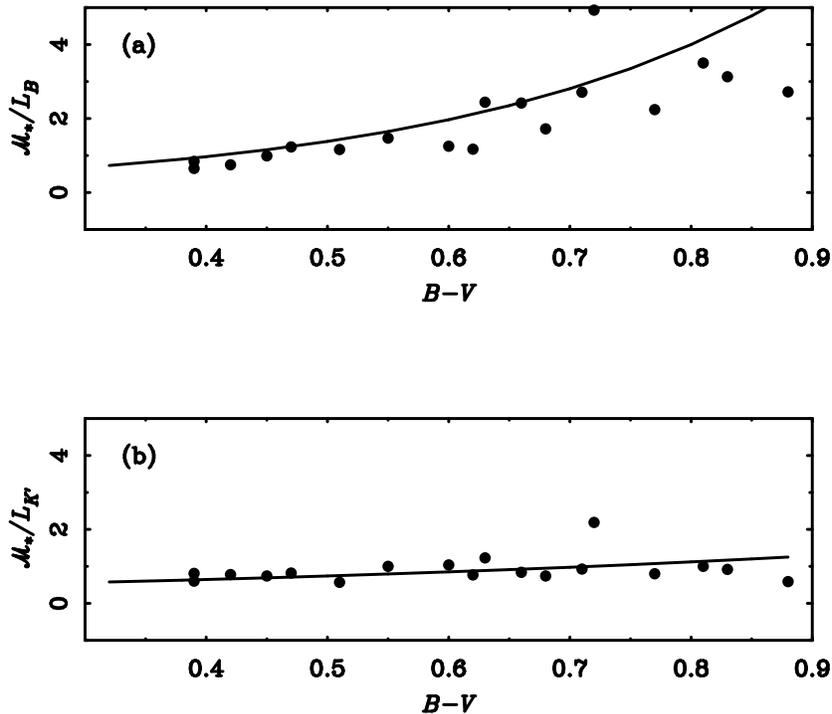}
\end{center}
\caption[]{The stellar mass-to-light ratios in (\textbf{a}) the $B$-band and
(\textbf{b}) the $K'$-band predicted by a slope 4 BTF for the UMa
galaxies \cite{Verh}, \cite{TVPHW}, \cite{VS}. 
These are plotted as a function of $B-V$
color, together with the Bruzual \& Charlot, Salpeter IMF model from 
\cite{BJ} (the first model in their table 4).  The population synthesis
models are in good agreement with the BTF, indicating that we have a
good handle on ${\cal M}_*/L$ and disk masses.}
\label{fig2}
\end{figure}

\section{Dark Matter Halos}

Rotation curves, by themselves, can only give a lower limit on the total
halo mass:  that enclosed by the last measured point.  However, if the
functional form of the halo were known, it might be possible to provide
some constraint by fitting the observations to the known form.
The NFW halo paradigm \cite{Moore},\cite{NFW}
which has arisen from cosmological N-body
simulations in principle gives a way to do this.  

Unfortunately, if not surprisingly, observed rotation curves never extend
far enough to constrain the circular velocity at the virial radius, $V_{200}$
\cite{MRB}.  There is a great deal of degeneracy between the concentration
$c$ and $V_{200}$.  An example is given in Fig.~3, which shows how
difficult it can be to distinguish between fits with NFW halos of rather
different parameters.

\begin{figure}[t]
\begin{center}
\includegraphics[width=.9\textwidth]{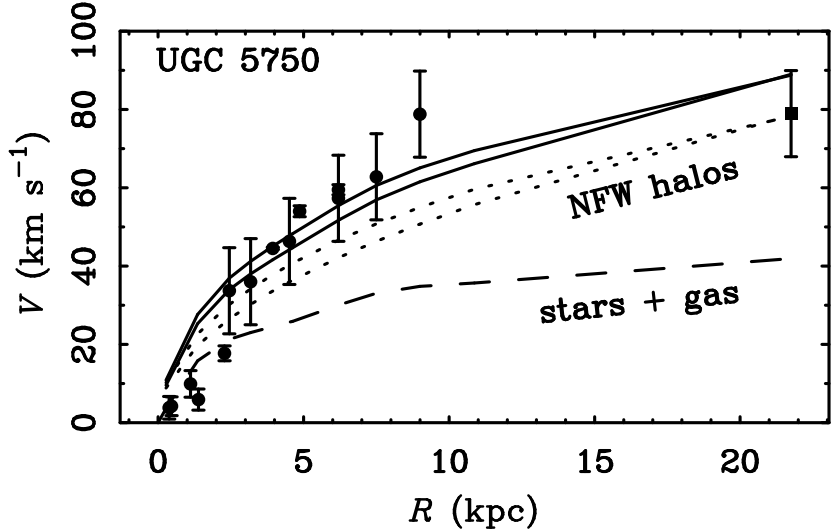}
\end{center}
\caption[]{An example of an NFW halo fit to the LSB galaxy UGC 5750 \cite{BMR}.
The best fit parameters
in this case are $c=1.9$ and $V_{200}=117$ for ${\cal M}_*/L_R = 1.4\;
{\cal M}_{\odot}/L_{\odot}$.  Another tolerable fit with
$c=0.2$ and $V_{200} = 300$ is also shown (lower dotted curve) to illustrate
the degeneracy between parameters.  Though many models sort of fit, their
concentrations are implausibly low for $\uLambda$CDM.
}
\label{fig3}
\end{figure}

Matters are made worse by the general failure of the NFW form to provide
a good description of the data.  The data just don't look like NFW halos.
Statistically, halos with constant density cores are almost always preferred
over those with cusps \cite{BMR}.
This is most clear in the best resolved cases \cite{BMBR}. 

The most important systematic concern at this point is not observational.
Resolution has improved by an order of magnitude \cite{SMT},
\cite{MRB}, \cite{BB} over the original 21 cm data for LSB galaxies
\cite{vdH}, \cite{BMH}.  The NFW shape has not become apparent as the data
have improved.  Instead, the systematics pointed out
by \cite{MB98a},\cite{MB98b} as problematic for CDM (independent
of the cusp issue) have only become more clear.  
Concern over slit mispositioning \cite{Swat} are misplaced:
independent observers reproduce one anothers' results \cite{MRB},\cite{BB}.
While there are certainly cases in which the error bars are large enough
to allow an NFW fit, isothermal fits are inevitably better.  Simply changing
the size of the error bars won't change this: a systematic change in the
shapes of $\sim 50$ high resolution rotation curves is required.  One can
certainly imagine ways in which this might happen \cite{Swat}, but it is
extremely unlikely that any of these ideas apply to real data, let
alone to \textbf{all} of the data from various independent sources.

The most serious issue is in the mass models:  stars have mass. 
Even in the limit of zero stellar mass, which is the
most favorable to the NFW case, isothermal halos are statistically
preferred \cite{BMR},\cite{BB}.  The situation
only becomes more grim if stars are allowed to have mass.  Though LSB
galaxies are dark matter dominated down to small radii, plausible ${\cal M}_*/L$
models do require that {\it some\/} of the velocity be attributed to luminous
mass.  This pulls the inferred dark matter distribution 
further away from the expected cusp slope.

We are hardly unique in reaching these conclusions, which are shared
by \textbf{all} published analyses of high resolution long slit H$\alpha$ data
\cite{BS},\cite{CCF},\cite{BB},\cite{BMBR},\cite{BMR},\cite{dong},\cite{Sal}.
High resolution Fabry-Perot \cite{BAC},\cite{Ben},\cite{PP} and CO \cite{BSLB}
data are also inconsistent with cuspy halos,
as are a variety of data for the Milky Way itself \cite{BE}.
The only analyses which are favorable to NFW are those
of low resolution data with large error bars \cite{BRDB},\cite{vdBS}.
When the error bars are large, any model can be driven through them.  
Though it has not been emphasized, constant density cores
provide as good or better fits even in these cases.

The isothermal halo form, while effective, is an extremely flexible
fitting function which
lacks the motivation of the NFW halo form.  So one might persists that the
NFW fits are still more appropriate in that they can be related to cosmology.
Standard $\uLambda$CDM makes a clear prediction \cite{NFW} for what the
concentrations of dark matter halos should be: $c = 9$ for $\Omega_m h = 0.2$.
Scatter about this value should be modest --- the largest estimate
\cite{Bullock} finds a lognormal distribution with $\sigma_c = 0.18$.
The median observed concentration is $c = 6.4$ \cite{BMR} which is different
from the standard $\uLambda$CDM prediction by many $\sigma$.  The problem
with NFW halos is not just a matter of getting fits to individual galaxies,
but also of understanding how the observed concentrations can be so low.
These low concentrations would be tolerable in a very low density
universe with $\Omega_m h \approx 0.12$ \cite{MBB}.  Even then there exists
a significant tail of very low concentration ($c < 4$) galaxies which simply
should not exist for any plausible cosmology.

The debate over halo profiles, while contentious, misses the real point.
Many halo profiles are nominally viable because they have lots of degenerate
free parameters.  Mass modeling is a bit like fitting a high order polynomial
to a few data points:  the line goes through the data, but means nothing.
One would prefer to have a minimal parameter description of the data.
Such a prescription exists \cite{IAP}.  It has long been noted that there
is a strong coupling between mass and light.\footnote{Renzo's Rule: when you
see a feature in the light, you see a corresponding feature in the rotation
curve.}  Oddly, this coupling persists for dark matter dominated LSB galaxies.
One needs only a single parameter per galaxy, the stellar mass-to-light ratio,
in order to fit the rotation curve in comparable or greater detail than
can be matched
by many-parameter halo models.  The mass-to-light ratio in the $K$-band is
close enough to constant that one can make a good zero parameter prediction
with such data \cite{SV}.  Until we come to terms with this observed
phenomenology, debating the cusp slope of dark matter halos is rather akin
to debating the number of angels that can dance on the head of a pin. \\

\noindent \textbf{Acknowledgements:} I am most grateful to Vera Rubin,
Erwin de Blok, and Albert Bosma for their work on the issues discussed here.
I would also like to thank Renzo Sancisi, Marc Verheijen, Rob Swaters,
and Frank van den Bosch for many lively and stimulating conversations.
No doubt, I have yet to hear the end of it!
The work of SSM is supported in part by NSF grant AST9901663.

%

\end{document}